\documentclass{iau} 
\usepackage{graphicx}
\usepackage{natbib}
\usepackage{hyperref}
\usepackage{wrapfig}

\title[New simulation of QSO X-ray heating during the CD] %% give here short title %%
{New simulation of QSO X-ray heating during the Cosmic Dawn}

\author[Hannah E. Ross, Keri Dixon, Ilian Iliev \& Garrelt Mellema]   %% give here short author list %%
{Hannah E. Ross$^{1*}$, Keri Dixon$^2$, Ilian Iliev$^1$,
 \and Garrelt Mellema$^4$}

\affiliation{$^1$Department of Physics and Astronomy, University of Sussex, Falmer, BN1 9QH, UK \\[\affilskip]
$^2$ New York University Abu Dhabi, Computational Research Building, Abu Dhabi,
UAE \\[\affilskip]
$^3$ AlbaNova SCFAB, Astronomi, 106 91, Stockholm, Sweden \\
$^*$Email: H.Ross@sussex.ac.uk}

\pubyear{2018}
\volume{333}  %% insert here IAU Symposium No.
\jname{Peering towards Cosmic Dawn}
\editors{Vibor Jeli\'c \& Thijs van der Hulst, eds.}
\begin{document}

\maketitle

\begin{abstract}
The upcoming radio interferometer Square Kilometre Array is expected to directly detect the redshifted 21-cm signal from the Cosmic Dawn for the first time. In this era temperature fluctuations from X-ray heating of the neutral intergalactic medium can impact this signal dramatically. Previously, in \citet{Ross2017}, we presented the first large-volume, 244~$h^{-1}$Mpc=349\,Mpc a side, fully numerical radiative transfer simulations of X-ray heating. This work is a follow-up where we now also consider QSO-like sources in addition to high mass X-ray binaries. Images of the two cases are clearly  distinguishable at SKA1-LOW resolution and have RMS fluctuations above the expected noise. The inclusion of QSOs leads to a dramatic increase in non-Gaussianity of the signal, as measured by the skewness and kurtosis of the 21-cm signal. We conclude that this increased non-Gaussianity is a promising signature of early QSOs. 
\end{abstract}

\firstsection % if your document starts with a section,
\section{Introduction}

The Epoch of Reionization (EoR) remains largely unconstrained with no direct observations having been made to date. The most promising observational probe of this time is the redshifted signal from the hypefine spin-flip transition of hydrogen (the 21-cm signal) which contains a vast amount of information about the EoR. Unlike its predecessors the anticipated Square Kilometer Array (SKA) will have the sensitivity required to detect this signal from the first stages of the EoR, known as the Cosmic Dawn (CD). 

During this period temperature variations in the still neutral IGM are thought to be one of the dominant contributors to the 21-cm fluctuations. The neutral IGM can only be heated by X-ray photons as they have long mean free paths, unlike the lower energy radiation emitted by stars. The 21-cm signal from this time is expected to be sensitive to the spectra, abundance and clustering of any X-ray sources present. 

The nature of these early X-ray sources remains uncertain. The first generation of stars (Pop III stars) could have formed binary systems as early as redshift 30. High mass X-ray binaries, HMXBs, are possibly significant contributors to early X-ray emissions \cite[e.g.][]{Xu2014,Jeon2014,Jeon2015}. In addition recent observations of high-redshift observations have suggested that QSOs may have been more abundant than previously thought. \citet{Giallongo2015,Bowler2012,Bowler2016,Stark2015a,Stark2015b,Stark2017}

We present a new full numerical simulation of inhomogeneous heating during the CD including QSO sources. Using multi-frequency radiative transfer (RT) modelling, we compare the morphology and evolution of the 21-cm signal to the results from \citep{Ross2017} and briefly discuss non-Gaussianity. The size of our simulations is sufficiently large to capture the large-scale patchiness.

\section{Methodology}

\subsection{Sources}

Haloes down to a mass of 10$^9\,$ M$_\odot$ were found with the spherical overdensity algorithm with an overdensity parameter of 178 with respect to the mean density. A sub-grid model \citep{Ahn2015a} calibrated with very high-resolution simulations to add the haloes down to $10^8\,$M$_\odot$, the minimum mass where the atomic line cooling of primordial gas is efficient.  The new simulation contains three types of source: stellar sources, HMXBs and QSOs. The details of these sources are outlined below. For further details see \cite{Ross2017}.

\subsubsection{Stellar sources}

Stellar sources form inside dark matter haloes with the luminosity proportional to the halo mass and live for 11.5 Mega-years. Sources hosted by high mass halos (HMACHs, 10$^9$M$_\odot$$\le$M) are unaffected by radiative feedback as their halo mass is above the Jeans mass for $\sim\!10^4$K. Halos with masses below the Jeans mass, but greater than the minimum mass at which atomic line cooling is efficient (LMACHs, 10$^8$M$_\odot$\textless M\textless 10$^9$M$_\odot$) have a higher efficiency factor than HMACHs due to the fact they likely contain more Pop.~III stars and are suppressible. Stellar sources have a blackbody spectrum with an effective temperature of $T_{\mathrm{eff}} = 5 \times 10^4 K$ (corresponding to the spectra of O and B type stars). 

\subsubsection{HMXBs}

The HMXB sources trace the stellar population and also have a luminosity proportional to the halo mass. HMXBs are assigned a power-law spectrum with an index of -1.5 in luminosity, extending from 272.08 eV to 100 times the second ionization of helium (5441.60 eV). The low frequency cut-off corresponds to the obscuration suggested to be present by observational works \citep[e.g.\ ][]{Lutivnov2005} and is consistent with the optical depth from high redshift gamma ray bursts \citep{Totani2006,Greiner2009}. For more details see \cite{Mesinger2013}. 

\subsubsection{QSO sources}

The X-ray emissivity from QSOs is quantified using the QLF from \citet{Ueda2014}. This QLF takes the form of a double power law with luminosity-density evolution and is best suited to an X-ray luminosity ($L_{\rm X}$) range of log($L_{\rm X}$) = 42 - 47.

The co-moving number density of QSOs, $n_{\rm q}$, is calculated by integrating the QLF, $\Phi(L,z)$ over all possible luminosities:
\begin{equation}
n_q = \int\limits_{L_{\rm min}}^{L_{\rm max}} \Phi (L,z) \  dL,
\end{equation}
where log($L_{\rm min}$) = 41, 41.5 for and log($L_{\rm max}$) = 47. The luminosity of the QSOs ($L_{\rm q}$) is then assigned by randomly sampling the QLF. The QSO spectrum is assumed to be:
\begin{equation}
L_{\mathrm{q}}(E) \propto E^{-\alpha_{\mathrm X}}
\end{equation} 
where $\alpha_{\mathrm X}$ = 0.8, from \citet{Ueda2014} via \citet{Haardt2015}. Like the HMXBs the spectra extends from 272.08eV to 5441.6eV. Unlike other source types, QSOs are not assumed to follow the density fluctuations, but instead are placed randomly in haloes larger than $10^9 M_\odot$. QSOs are active for $3 \times 11.5$Myr timesteps, which is consistent with current estimates \citep[e.g.][]{Borisova2016,Khrykin2016}. A every timestep the QSOs are assigned a new $L_{\rm q}$ within an order of magnitude of the original $L_{\rm q}$ to mimic the variability observed QSOs. If the host halo of a QSO moves the QSO is placed in the nearest halo to its previous location in order to track it. This effect is not significant, only occurring for ten out of tens of the thousands of QSOs present in the simulation volume. At early times in the simulation there are not enough HMACHs to host sufficient numbers of QSOs to reproduce the luminosity function. At these times we assign QSOs to each available halo and discard the remaining QSOs.

\subsection{The simulations}
\label{sec:sims}

The density fields and halo catalogues were obtained from a high-resolution $N$-body simulation (presented in \citet{Dixon2016} and completed under the Partnership for Advanced Computing in Europe, PRACE, Tier-0 project called PRACE4LOFAR). This simulation was run using the \textsc{\small CubeP$^3$M} code \citep{Harnois2013}. It followed $4000^3$ particles in a $349\,Mpc$ per side volume to enable halo identification of the smallest HMACHs. 

The radiative transfer simulation was run using the latest version of the photon conserving, short characteristics ray-tracing method \textsc{\small C$^2$-Ray} Code \citep{Mellema2006,Friedrich2012}. The boxsize is $349\,Mpc$ and grid-size of 250$^3$. The simulation was performed under the PRACE Tier-0 projects PRACE4LOFAR and Multi-scale Reionization.

\begin{figure*} 
\centering
\makebox[\textwidth][c]{\includegraphics[width=1.\textwidth]{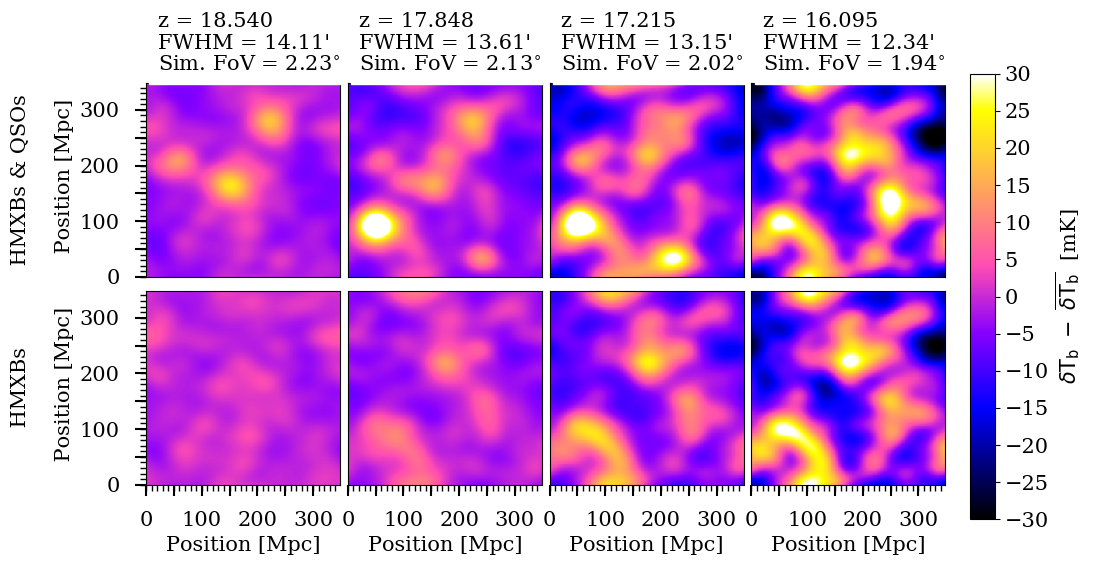}
}
\caption{Mean-subtracted differential brightness temperature maps smoothed with a Gaussian beam with the full-width half-maximum (FWHM) corresponding to a 1.2~km maximum baseline at the relevant frequency, as labelled. The images are bandwidth-smoothed with a top hat function (width equal to the distance corresponding to the beam width). The simulation with HMXBs and QSOs runs along the top row and the previously presented HMXB case is below, with snapshots of the same redshifts being vertically aligned.}
\label{fig:dbtmaps}
\end{figure*}

\section{Results and Discussion}

The observable quantity is the differential brightness temperature ($\delta T_\mathrm{b}$) which is seen in absorption or emission with respect to the CMB. As SKA will not be able to detect the absolute value of $\delta T_\mathrm{b}$, it is prudent to focus on the fluctuations of the signal. During the CD temperature variations are able to introduce fluctuations into the 21-cm signal until the temperature is much greater than that of the CMB, known as temperature saturation (see \citet{Ross2017}). 

\begin{wrapfigure}{r}{0.4\textwidth}
  \begin{center}
    \includegraphics[width=0.4\textwidth]{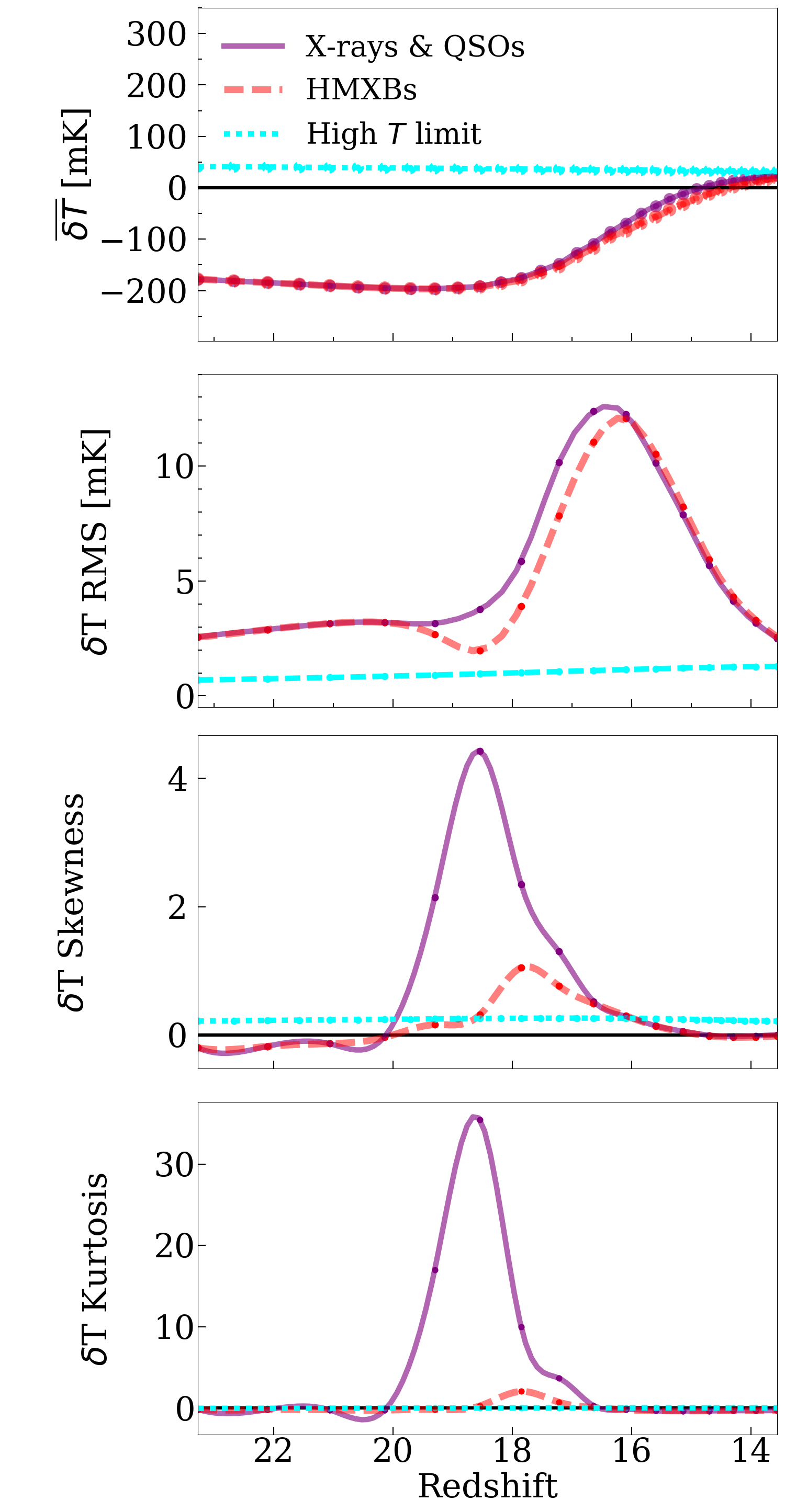}
  \end{center}
  \caption{Statistics from the 21-cm signal from both simulations as well as the high-$T$$_\mathrm{K}$ limit are shown. The top left panel shows the mean value of $\delta T_{\rm b}$, the bottom left panel the rms, the top right panel the skewness and the bottom right the kurtosis.}
  \label{fig:statistics}
\end{wrapfigure}

Maps of the $\delta T_\mathrm{b}$ at the expected SKA1-LOW resolution from both heating models are displayed in Fig. \ref{fig:dbtmaps} for selected redshifts. In the angular direction coeval cubes are smoothed with a Gaussian beam with a full-width half-maximum corresponding to a 1.2~km maximum baseline at the relevant frequency. Cubes are bandwidth-smoothed with a top hat function (with width corresponding to the same physical distance as the FWHM of the beam).
 
The two models are clearly distinguishable even at the SKA1-Low resolution, as visible in Fig \ref{fig:dbtmaps}. In our model the heated regions of individual QSOs are distinguishable at the higher redshifts. As the transition to emission is approached the heating fluctuations have a smaller effect, meaning that although there are more QSOs at later times, the early QSOs are the most detectable.

In the top panel of Fig. \ref{fig:statistics} the mean $\delta T_\mathrm{b}$ is shown along with the High-$T_\mathrm{K}$. Both models reach temperature saturation at similar times, with the HMXB + QSO model only marginally before (at z=13.557 rather than 12.459). The evolution of mean is comparable between the two simulations, with marginally higher values in the HMXB + QSO case. In these models temperature saturation is reached while the number of QSOs is small, meaning that the QSOs have not been able to contribute sufficient energy to impact the overall heating of the volume before temperature saturation occurs. Hence the mean values remain similar.

In the second panel from the top of Fig. \ref{fig:statistics} the rms fluctuations of $\delta T_\mathrm{b}$ are shown for both cases. Again, the value for the new case with both types of source is marginally higher as QSOs are able to introduce additional heating to regions where HMXBs are not particularly bright. However, due to the low number of QSOs, the effect is small. Fluctuations from both models are well above the expected noise for deep integrations with the SKA1-Low, indicating that the detection of the X-ray heating epoch could also be feasible for the case of X-ray heating by both HMXBs and QSOs.

Unlike the lower order statistics, the skewness and kurtosis (see the lower two panels of Fig \ref{fig:statistics}) show an extremely different evolution. The skewness is four times greater in the HMXBs and QSO case than the original case, and the kurtosis is an order of magnitude greater. From this we conclude that even in these small numbers QSOs are able to introduce significant non-Gaussianity to the signal. This is due to the fact that, unlike HMXBs, the QSOs do not have luminosities proportional to their host halos. In the case of HMXBs the amount of heating in each region is proportional to the mass present there, so although the signal is largely non-Gaussian, it still has some correlation to the Gaussian underlying cosmic structures. However, when QSOs are added to the simulation their luminosities are independent of the halo mass, which introduces variations uncorrelated to the density fluctuations. These results suggest that tests of non-Gaussianity could be a useful probe for QSO like sources at this time.


\begin{thebibliography}{}

\bibitem[Ahn \etal\ (2015)]{Ahn2015a}
{{Ahn}, K. and {Xu}, H. and {Norman}, M.~L.} 2015,
\textit{Astronomy and Astrophysics}, 802, 8

\bibitem[Borisova \etal\ (2016)]{Borisova2016}
{{Borisova}, E. and {Lilly}, S.~J. and {Cantalupo}, S. et al} 2016,
\textit{The Astrophysics Journal}, 830, 120

\bibitem[Bowler \etal\ (2012)]{Bowler2012}
{{Bowler}, R.~A.~A. and {Dunlop}, J.~S. and {McLure}, R.~J.et al} 2012,
\textit{Monthly Notices of the Royal Astronomical Society}, 426, 2772-2788

\bibitem[Bowler \etal\ (2016)]{Bowler2016}
{{Bowler}, R.~A.~A. and {Dunlop}, J.~S. and {McLure}, R.~J.et al} 2016,
\textit{Monthly Notices of the Royal Astronomical Society}, 452, 1817-1840

\bibitem[Dixon \etal\ (2016)]{Dixon2016}
{{Dixon}, K.~L. and {Iliev}, I.~T. and {Mellema}, G. et al} 2016,
\textit{Monthly Notices of the Royal Astronomical Society}, 456, 3011-3029

\bibitem[Friedrich \etal\ (2012)]{Friedrich2012}
{{Friedrich}, M.~M. and {Mellema}, G. and {Iliev}, I.~T. and 
	{Shapiro}, P.~R.} 2012,
\textit{Monthly Notices of the Royal Astronomical Society}, 421, 2232-2250

\bibitem[Giallongo \etal\ (2015)]{Giallongo2015}
{{Giallongo}, E., {Grazian}, A., {Fiore}, F. et al} 2015,
\textit{Astronomy and Astrophysics}, 587, A83 

\bibitem[Greiner \etal\ (2009)]{Greiner2009}
{{Greiner}, J. and {Kr{\"u}hler}, T. and {Fynbo}, J.~P.~U. et al} 2009,
\textit{Astronomy and Astrophysics}, 693, 1610-1620 

\bibitem[Haardt \etal\ (2015)]{Haardt2015}
{{Haardt}, F. and {Salvaterra}, R.} 2015,
\textit{Astronomy and Astrophysics}, 575, L16

\bibitem[Harnois-D{\'e}raps \etal\ (2013)]{Harnois2013}
{{Harnois-D{\'e}raps}, J. and {Pen}, U.-L. and {Iliev}, I.~T. et al} 2013,
\textit{Monthly Notices of the Royal Astronomical Society}, 436, 540-559

\bibitem[Jeon \etal\ (2014)]{Jeon2014}
{{Jeon}, M. and {Pawlik}, A.~H. and {Bromm}, V. and {Milosavljevi{\'c}}, M.} 2014,
\textit{Monthly Notices of the Royal Astronomical Society}, 440, 3778-3796 

\bibitem[Jeon \etal\ (2015)]{Jeon2015}
{{Chen}, K.-J. and {Bromm}, V. and {Heger}, A. and {Jeon}, M. and 
	{Woosley}, S.} 2015,
\textit{Monthly Notices of the Royal Astronomical Society}, 802, 13 

\bibitem[Jeon \etal\ (2015)]{Jeon2015}
{{Chen}, K.-J. and {Bromm}, V. and {Heger}, A. and {Jeon}, M. and 
	{Woosley}, S.} 2015,
\textit{Monthly Notices of the Royal Astronomical Society}, 802, 13 

\bibitem[Khrykin \etal\ (2016)]{Khrykin2016}
{{Khrykin}, I.~S. and {Hennawi}, J.~F. and {McQuinn}, M.} 2016,
\textit{The Astrophysics Journal}, 838, 96

\bibitem[Lutivnov \etal\ (2005)]{Lutivnov2005}
{{Lutovinov}, A. and {Revnivtsev}, M. and {Gilfanov}, M. et al} 2005,
\textit{Astronomy and Astrophysics}, 444, 821-829

\bibitem[Mellema \etal\ (2006)]{Mellema2006}
{{Mellema}, G. and {Iliev}, I.~T. and {Alvarez}, M.~A. and {Shapiro}, P.~R.} 2006,
\textit{Monthly Notices of the Royal Astronomical Society}, 11, 374-395

\bibitem[Mesinger \etal\ (2013)]{Mesinger2013}
{{Mesinger}, A. and {Ferrara}, A. and {Spiegel}, D.~S.} 2013,
\textit{Monthly Notices of the Royal Astronomical Society}, 431, 621-637

\bibitem[Ross \etal\ (2017)]{Ross2017}
{{Ross}, H.~E. and {Dixon}, K.~L. and {Iliev}, I.~T. and {Mellema}, G.} 2017,
\textit{Monthly Notices of the Royal Astronomical Society}, 468, 3785-3797

\bibitem[Stark \etal\ (2015a)]{Stark2015a}
{{Stark}, D.~P. and {Richard}, J. and {Charlot}, S. et al} 2015,
\textit{Monthly Notices of the Royal Astronomical Society}, 450, 1846-1855

\bibitem[Stark \etal\ (2015b)]{Stark2015b}
{{Stark}, D.~P. and {Walth}, G. and {Charlot}, S. and {Cl{\'e}ment}, B. et al} 2015,
\textit{Monthly Notices of the Royal Astronomical Society}, 454, 1393-1403

\bibitem[Stark \etal\ (2017)]{Stark2017}
{{Stark}, D.~P. and {Ellis}, R.~S. and {Charlot}, S. and {Chevallard}, J. et al} 2017,
\textit{Monthly Notices of the Royal Astronomical Society}, 464, 469-479

\bibitem[Totani \etal\ (2006)]{Totani2006}
{{Totani}, T. and {Kawai}, N. and {Kosugi}, G. et al} 2006,
\textit{IAU Joint Discussion}, 7

\bibitem[Ueda \etal\ (2014)]{Ueda2014}
{{Ueda}, Y. and {Akiyama}, M. and {Hasinger}, G. and {Miyaji}, T. and 
	{Watson}, M.~G.} 2014,
\textit{The Astrophysics Journal}, 786, 104

\bibitem[Xu \etal\ (2014)]{Xu2014}
{{Xu}, H., {Ahn}, K., {Wise}, J.~H.,{Norman}, M.~L. and 
	{O'Shea}, B.~W.} 2014,
\textit{The Astrophysics Journal}, 791, 110 


\end{thebibliography}
\end{document}